# SPALLATION NEUTRON SOURCE AND OTHER HIGH INTENSITY PROTON SOURCES[*]

WEIREN CHOU

*Fermi National Accelerator Laboratory*
*P.O. Box 500*
*Batavia, IL 60510, USA*
*E-mail: chou@fnal.gov*

This lecture is an introduction to the design of a spallation neutron source and other high intensity proton sources. It discusses two different approaches: linac-based and synchrotron-based. The requirements and design concepts of each approach are presented. The advantages and disadvantages are compared. A brief review of existing machines and those under construction and proposed is also given. An R&D program is included in an appendix.

## 1. Introduction

### 1.1. *What is a Spallation Neutron Source?*

A spallation neutron source is an accelerator-based facility that produces pulsed neutron beams by bombarding a target with intense proton beams.

Intense neutrons can also be obtained from nuclear reactors. However, the international nuclear non-proliferation treaty prohibits civilian use of highly enriched uranium $U^{235}$. It is a showstopper of any high efficiency reactor-based new neutron sources, which would require the use of 93% $U^{235}$. (This explains why the original proposal of a reactor-based Advanced Neutron Source at the Oak Ridge National Laboratory in the U.S. was rejected. It was replaced by the accelerator-based Spallation Neutron Source, or SNS, project.)

A reactor-based neutron source produces steady higher flux neutron beams, whereas an accelerator-based one produces pulsed lower flux neutron beams. So the trade-off is high flux *vs.* time structure of the neutron beams. This course will teach accelerator-based neutron sources.

An accelerator-based neutron source consists of five parts:
1) Accelerators
2) Targets
3) Beam lines

---

[*] This work is supported by the Universities Research Association, Inc., under contract No. DE-AC02-76CH03000 with the U.S. Department of Energy.

4) Detectors
5) Civil construction

A project proposal includes 1) through 5), plus a cost estimate, a schedule and environment, safety and health (ES&H) considerations. This course will teach part 1) only, although part 2) is closely related to 1) and a critical item in the design of a spallation neutron source.

### 1.2. *Parameter Choice of a Spallation Neutron Source*

The requirements of neutron beams for neutron scattering experiments are as follows:
- Neutron energy: low, about a few milli electron volts.
- Neutron pulse: sharp, about 1 µs.
- Pulse repetition rate: 10-60 Hz.

When an intense proton beam strikes on a target (made of carbon or heavy metal), neutrons are produced via spallation. The production rate is roughly proportional to the power deposited on the target.

Proton energies between 1 and 5 GeV prove optimal for neutron production. At 1 GeV, each incident proton generates 20-30 neutrons.

The beam power $P$ is the product of beam energy $E$, beam intensity $N$ (number of protons per pulse) and repetition rate $f$:

$$P \text{ (MW)} = 1.6 \times 10^{-16} \times E \text{ (GeV)} \times N \times f \text{ (Hz)} \qquad (1)$$

Typical parameters of a modern high power spallation neutron source are:
- $P \sim 1$ MW
- $E \sim 1$ GeV
- $N \sim 1 \times 10^{14}$
- $f \sim 10\text{-}60$ Hz

### 1.3. *Linac-based vs. Synchrotron-based Spallation Neutron Source*

There are two approaches to an accelerator-based spallation neutron source: linac-based and synchrotron-based.

A linac-based spallation neutron source has a full-energy linac and an accumulator ring. It works as follows:
- A heavy-duty ion source generates high intensity H$^-$ beams.
- A linac accelerates H$^-$ pulses of ~1 ms length to ~1 GeV.
- These H$^-$ particles are injected into an accumulator via a charge exchange process, in which the electrons are stripped by a foil and dumped, and the H$^+$ (proton) particles stay in the ring.

- This injection process takes many (several hundreds to a few thousands) turns.
- The accumulated protons are then extracted from the ring in a single turn onto a target. The pulse length is about 1 μs.
- This process repeats 10-60 times every second.

A synchrotron-based spallation neutron source has a lower energy linac and a rapid cycling synchrotron. It works differently.

- A heavy-duty ion source generates high intensity H$^-$ beams.
- A linac accelerates H$^-$ pulses of ~1 ms length to a fraction of a GeV.
- These H$^-$ particles are injected into a synchrotron via the same charge exchange process.
- This injection process takes many (several hundreds) turns.
- The H$^+$ (proton) beam is accelerated in the synchrotron to 1 GeV or higher and then extracted in a single turn onto a target. The pulse length is about 1 μs.
- This process repeats 10-60 times every second.

Compared with a linac-based spallation neutron source, a synchrotron-based one has the following advantages:

- For the same beam power, it would cost less, because proton synchrotrons are usually less expensive than proton linacs.
- For the same beam power, it would have lower beam intensity, because the beam energy could be higher.
- Because the injected linac beam has lower power, the stripping foil is easier. Also, larger beam loss at injection could be tolerated.
- A major problem in a high intensity accumulator ring (a DC machine) is the e-p instability. However, this has never been observed in any synchrotron (an AC machine) during ramp.

The disadvantages of a synchrotron-based spallation neutron source include:

- AC machines (synchrotrons) are more difficult to build than DC machines (accumulators). In particular, the hardware is challenging, e.g., large aperture AC magnets, rapid cycling power supplies, field tracking during the cycle, eddy current effects in the coil and beam pipe, high power tunable RF system, etc.
- AC machines are also more difficult to operate than DC machines. Therefore, the reliability is lower.

One needs to consider all these factors when deciding which approach to take for a spallation neutron source.

**1.4.** *Spallation Neutron Source vs. Other High Intensity Proton Sources*

Spallation neutron sources are an important type of high intensity proton sources. However, a high intensity proton source may find many other applications. For example:
- To generate high intensity secondary particles for high-energy physics experiments, e.g., antiprotons (Tevatron p-pbar collider), muons (AGS, JHF), neutrinos (K2K, MiniBooNE, NuMI, JHF), kaons (CKM, KAMI, JHF), ions (ISOLDE), etc.
- To generate neutrino superbeams as the first stage to a neutrino factory and a muon collider. (Such a high intensity proton source is called a Proton Driver.)
- Nuclear waste transmutation (JHF, CONCERT).
- Energy amplifier (CERN).
- Proton radiography (AHF).

The design concept learned from this course can readily be applied to the design of other high intensity proton sources.

**2. High Intensity Proton Sources: Existing, Under Construction, and Proposed**

There are a number of high intensity proton sources operating at various laboratories over the world. There are presently two large construction projects: the SNS at the Oak Ridge National Laboratory in the U.S., and the JHF at the KEK/JAERI in Japan. Each has a construction budget of about 1.3 billion US dollars and is scheduled to start operation around 2005-2006. There are also numerous proposals for Proton Drivers and other high intensity proton sources. Table 1 is a summary based on a survey conducted during the Snowmass 2001 Workshop.[1]

Among the existing machines, the highest beam power from a synchrotron is 160 kW at the ISIS at Rutherford Appleton Laboratory in England. The highest beam power from an accumulator is 64 kW at the PSR at Los Alamos National Laboratory in the U.S.

The SNS is a linac-based spallation neutron source. The design beam energy is 1 GeV, beam power 1.4 MW. The JHF is a synchrotron-based facility. It has a 400 MeV linac, a 3 GeV rapid cycling synchrotron with a beam power of 1 MW, and a 50 GeV slow ramp synchrotron with a beam power of 0.75 MW.

Several proposals of proton drivers have been documented and can be found in Ref. [2]-[4].

Table 1. High intensity proton sources: existing, under construction, and proposed
(Snowmass 2001 survey)

| Machine | Flux ($10^{13}$/pulse) | Rep Rate (Hz) | Flux† ($10^{20}$/year) | Energy (GeV) | Power (MW) |
|---|---|---|---|---|---|
| **Existing:** | | | | | |
| RAL ISIS | 2.5 | 50 | 125 | 0.8 | 0.16 |
| BNL AGS | 7 | 0.5 | 3.5 | 24 | 0.13 |
| LANL PSR | 2.5 | 20 | 50 | 0.8 | 0.064 |
| ANL IPNS | 0.3 | 30 | 9 | 0.45 | 0.0065 |
| Fermilab Booster (*) | 0.5 | 7.5 | 3.8 | 8 | 0.05 |
| Fermilab MI | 3 | 0.54 | 1.6 | 120 | 0.3 |
| CERN SPS | 4.8 | 0.17 | 0.8 | 400 | 0.5 |
| **Under Construction:** | | | | | |
| ORNL SNS | 14 | 60 | 840 | 1 | 1.4 |
| JHF 50 GeV | 32 | 0.3 | 10 | 50 | 0.75 |
| JHF 3 GeV | 8 | 25 | 200 | 3 | 1 |
| **Proton Driver Proposals:** | | | | | |
| Fermilab 8 GeV | 2.5 | 15 | 38 | 8 | 0.5 |
| Fermilab 16 GeV | 10 | 15 | 150 | 16 | 4 |
| Fermilab MI Upgrade | 15 | 0.65 | 9.8 | 120 | 1.9 |
| BNL Phase I | 10 | 2.5 | 25 | 24 | 1 |
| BNL Phase II | 20 | 5 | 100 | 24 | 4 |
| CERN SPL | 23 | 50 | 1100 | 2.2 | 4 |
| RAL 15 GeV (**) | 6.6 | 25 | 165 | 15 | 4 |
| RAL 5 GeV (**) | 10 | 50 | 500 | 5 | 4 |
| **Other Proposals:** | | | | | |
| Europe ESS (**) | 46.8 | 50 | 2340 | 1.334 | 5 |
| Europe CONCERT | 234 | 50 | 12000 | 1.334 | 25 |
| LANL AAA | - | CW | 62500 | 1 | 100 |
| LANL AHF | 3 | 0.04 | 0.03 | 50 | 0.003 |
| KOMAC | - | CW | 12500 | 1 | 20 |
| CSNS/Beijing | 1.56 | 25 | 39 | 1.6 | 0.1 |

† 1 year = $1 \times 10^7$ seconds.
(*) Including planned improvements.
(**) Based on 2-ring design.

## 3. Design Concept of a Linac-based Spallation Neutron Source

A linac-based spallation neutron source has three major accelerator components:
- Linac front end
- Linac (full energy)
- Accumulator

We will discuss the design concept of each component in the following sections.

### 3.1. *Linac Front End*

The linac front end consists of an ion (H$^-$) source, a pre-accelerator (Cockcroft-Walton or RFQ), a low energy beam transport (LEBT), and a chopper.

#### 3.1.1. *H$^-$ source*

H$^-$ ions have almost been universally adopted for multi-turn injection from a linac to an accumulator ring. These ions are generated in an H$^-$ source. There are several different types: surface-plasma source (magnetron), semi-planatron, surface-plasma source with Penning discharge (Dudnikov-type source), RF volume source, etc. This is a highly specialized field. There are regular conferences and workshops devoted to this topic. The main challenges are to provide ion beams with high brightness (i.e., high intensity and low emittance) and to operate at high duty factor with a reasonable lifetime.

#### 3.1.2. *Cockcroft-Walton and RFQ*

The kinetic energy of the H$^-$ particles from an ion source is about a few tens of keV. These particles are accelerated by a pre-accelerator, which can be either a Cockcroft-Walton or RFQ. The former has been in use for many years and has a maximum energy of about 750 keV. In a number of laboratories it has been replaced by the latter, which is a common choice of new accelerators. This is because an RFQ has higher energy (several MeV) and a much smaller physical size. Its beam has higher brightness and is bunched. (The Cockcroft-Walton needs a buncher.) The design issues of an RFQ include high beam current, high efficiency, small emittance dilution, and higher order mode (HOM) suppression.

#### 3.1.3. *LEBT*

When an RFQ is used, one needs a low energy beam transport (LEBT) as a matching section between the ion source and the RFQ. It consists of lenses that focus the beam from the ion source, which is relatively large in radius and divergence. The LEBT also usually contains source diagnostics and provides the differential vacuum pumping between the source and the RFQ.

#### 3.1.4. *Chopper*

The purpose of a chopper is to chop the beam so that it can properly fit into the RF bucket structure in an accumulator. This would greatly reduce the injection loss caused by RF capture. The requirements of a chopper are: short rise- and fall-time (10-20 ns), short physical length (to reduce space charge effects), and a flat top and a flat bottom in the field waveform (to reduce energy spread in the beam).

There are several different types of choppers:
- Transverse deflector: This is a traveling wave structure. It has short rise- and fall-time. The shortcoming is its physical size (about 1-meter long). It is used at the Los Alamos National Laboratory and the Brookhaven National Laboratory.
- Electric deflector: This is a split-electrode structure for deflecting the beam right after the LEBT. It was built at the Lawrence Berkeley National Laboratory and will be installed in the linac front end of the SNS project. [5]
- Beam transformer (energy chopper): This is a new type of chopper and is based on the fact that an RFQ has a rather small energy window. A pulsed beam transformer that provides 10% energy modulation to the beam in front of an RFQ can effectively chop the beam. It has short rise- and fall-time and a short physical length (about 10 cm). A prototype has been built by a KEK-Fermilab team and is installed at the HIMAC in Japan for beam testing. [6]

## 3.2. *Linac*

The linac is the main accelerator. Its function is to accelerate the H$^-$ beam to full energy (~ 1 GeV) before injection into the accumulator. Because the particle velocity changes over a wide range during the acceleration ($\beta$ = 0.046 at 1 MeV, $\beta$ = 0.875 at 1 GeV), the linac is partitioned to several parts. Each part uses a different design to best match the corresponding $\beta$ values.

### 3.2.1. *Low energy part (below 100 MeV, $\beta$ < 0.4)*

Drift tube linac (DTL) is a common choice of this part. It is a matured technology and has been used in every proton linac over the world. A potential concern is that some vacuum tubes used to drive the RF cavities could have supply problem because the vendors may terminate their production.

There is also an effort to develop superconducting RF cavities (the so-called spoke cavity) for low $\beta$ acceleration.

### 3.2.2. *Medium energy part (100 MeV - 1 GeV, 0.4 < $\beta$ < 0.9)*

This is the bulk part of the linac. There are two design choices. One is room temperature coupled-cell linac (CCL), another superconducting (SC) linac.

The CCL is a matured technology and has been used in all existing linacs (e.g., Fermilab, Los Alamos National Laboratory, Brookhaven National Laboratory, etc.). The highest energy using this technology reaches 800 MeV (LANSCE at Los Alamos). However, the new project SNS has decided to use an SC linac for good reason.

SC linacs have been proved reliable and efficient in electron machines (e.g., LEP and CEBAF). But still, it is a challenge for employment in proton machines when operating in short pulse mode and accelerating particles with different $\beta$ values. In the past decade, SC linac technology has been making good and steady progress. [7] Compared with a room temperature linac, an SC linac has the following advantages:

- Higher accelerating gradient.
- Larger aperture (which is particularly important for high intensity beams).
- Lower operation cost.
- Lower capital cost if higher energy is required. (There is an energy threshold above which an SC linac becomes more economical.)

In addition to the SNS, CERN and KEK are also planning to use an SC linac in their future machines (SPL and JHF Stage 2 linac, respectively).

Among various challenges to an SC linac, a crucial one is the RF control. The allowable phase error ($< 0.5°$) and amplitude error ($< 0.5\%$) are demanding. One needs to investigate the choice of RF source (number of cavities per klystron), redundancy (off-normal operation with missing cavities), feedback and feedforward technique.

3.2.3. *High energy part (above 1 GeV, $\beta > 0.9$)*

In this range, particles travel at a velocity near that of the light and behave similar to electrons. An SC linac is an obvious and probably also the only choice from economical considerations. Several new high-energy proton linac proposals (2.2 GeV at CERN, 3 GeV at Los Alamos, and 8 GeV at Fermilab) have all picked this design.

**3.3.** *Accumulator*

As the name indicates, an accumulator is a ring that accumulates many turns of injected particles and ejects them in a single turn. The purpose is to convert long beam pulses (~ 1 ms) to short beam pulses (~ 1 μs) for experiments. It is a DC machine. Its hardware is more or less straightforward (a main advantage of the accumulator approach). But this by no means implies an *"easy"* machine. On the contrary, there are a number of challenges due to high beam power.

### 3.3.1. *Beam loss control*

This is the most challenging problem. For a 1 MW beam power, 1% beam loss would give 10 kW, which already exceeds the full beam power on the targets for most of the existing physics experiments. Therefore, allowable beam loss must be much lower than 1%.

There are two types of beam loss: controllable and uncontrollable. One uses specially designed collimators and dumps to collect the former so that the loss can be localized. The uncontrollable beam loss would spread over the entire machine and must be kept very low. The rule of thumb is that it must be below 1 W/m in order to make hands-on maintenance possible. For a 100-meter machine, 1 W/m gives the total uncontrollable beam loss of 100 W, which is 0.01% of the total beam power. This is a goal not impossible but very difficult.

In the PSR at Los Alamos, which is a 64 kW accumulator, the total beam loss is a fraction of a percent. Most of them are unstripped $H^0$ and $H^-$ particles, which are collected by special beam dumps.

### 3.3.2. *Collimators and remote handling*

Collimators are a critical part of an accumulator. They are used to localize the beam loss and leave a majority part of the machine "clean."

Modern collimators use a 2-stage design. The primary collimator scatters the halo particles; the secondary collimator (which can be more than one) collects them. There is one set of collimators in each transverse plane. Longitudinal collimators are also used, which are placed in high dispersion regions. The design efficiency of collimators is 95% or higher.

The area near the collimators is very "hot" (highly radioactive). One must use remote handling for maintenance in this area. Robot arms and cranes are often employed. This should be an integral part in the machine design. Invaluable experiences can be learned from LANSCE (Los Alamos, U.S.) and PSI (Switzerland). These machines have been handling MW beams for years and have designed several remote-handling systems that work reliably. [8]

### 3.3.3. *$H^-$ injection*

This is another difficult part of the design and has many technical issues involved.

- The stripping foil (usually made of carbon) must stand for high temperature and large shock waves. It must also have high stripping efficiency and a reasonable lifetime.
- The stripped electrons and unstripped $H^0$ and $H^-$ particles must be collected.
- During the many-turn (hundreds to thousands) injection, the orbit bump needs to "paint" the particles in the phase space so that a uniform distribution can be obtained. This would reduce the space charge effect. The orbit bump also needs to minimize the average number of hits per particle on the foil.
- The beam emittance dilution due to Coulomb scattering from the foil should be kept under control.
- There are proposals from the KEK and Los Alamos for laser stripping. The R&D is being pursued.

3.3.4. *Lattice*

The main requirement is multiple long straight sections, which are used for, respectively, injection, extraction, RF and collimation.

3.3.5. *e-p instability*

This is a main beam dynamics problem in an accumulator. In the PSR at Los Alamos, e-p instability is the bottleneck limiting the beam power. When the beam intensity reaches a threshold, rapid beam oscillations (usually in one transverse plane) occur that leads to fast beam loss. This instability is believed to be caused by electrons trapped in the proton bunch gap. These electrons come mainly from secondary yield. When a primary electron hits the wall, secondary electrons are generated, which are accelerated by the proton beam and hit the opposite side of the wall, generating more electrons, so on and so forth, causing an avalanche.

This so-called electron cloud effect (ECE) has also been seen in electron storage rings, in particular in the two B-factories: PEP-II and KEK-B. These two machines effectively used solenoids to suppress this effect. However, solenoids or clearing electrodes appear to be less useful in the PSR. (This is a puzzle to be resolved.) Instead, the following measures have been found effective in raising the instability threshold in the PSR: beam scrubbing (which conditions the wall), inductive inserts (which make the proton bunch gap cleaner), and sextupoles (which couple the motion in the two transverse planes and stabilize the oscillation in one plane).

This is an active research field. [9]

### 3.3.6. *Hardware*

The challenge is the magnet, which must have large aperture in order to accommodate large beam size and beam halo. Good field quality is necessary to ensure large dynamic aperture.

Other technical systems, including power supplies, RF, vacuum and diagnostics, are relatively straightforward.

## 4. Design Concept of a Synchrotron-based Spallation Neutron Source

A synchrotron-based spallation neutron source also has three major accelerator components:
- Linac front end
- Linac (lower energy)
- Synchrotron

The designs of the linac front end and the linac are similar to that of a linac-based spallation neutron source. However, a synchrotron design is very different from an accumulator. Therefore, we will focus on the synchrotron in this section.

### 4.1. *Lattice*

There are two basic requirements on the design: a transition-free lattice, and several dispersion-free straight sections. For high intensity operation in proton synchrotrons, transition crossing is often a major cause of beam loss and emittance blowup. One should avoid it in the first place. Dispersion in the RF, which is placed in one or more straight sections, may lead to synchro-betatron coupling resonance and should also be avoided.

For a medium energy synchrotron (above ~ 6 GeV), regular FODO lattices (in which $\gamma_t \propto \sqrt{R}$, $\gamma_t$ the lattice transition $\gamma$, $R$ the machine radius) are ruled out because they would use too many cells to achieve a high $\gamma_t$. Otherwise a transition crossing is inevitable when the $\gamma$ of the beam approaches $\gamma_t$ during ramp. There are several lattices that can give either a high or an imaginary $\gamma_t$ so that a transition crossing would not occur. For example, (a) a flexible momentum compaction (FMC) lattice, which has a singlet 3-cell modular structure with a missing or short dipole in the mid-cell; (b) a doublet 3-cell modular structure with a missing or short dipole in the mid-cell. Figure 1 is an example of (b), which is designed for a new 8 GeV synchrotron at Fermilab.

The choice of phase advance per module is of critical importance in this type of lattice. There are two reasons. (i) The chromaticity sextupoles are placed in the mid-cell, where the beta-function peaks and available space exists. In order to cancel the higher order effects of these sextupoles, they need to be paired properly. (ii) The phase advance per arc in the horizontal plane must be multiple of $2\pi$ in order to get zero dispersion in the straights without using dispersion suppressors (which are space consuming). Other requirements in the lattice design include: ample space for correctors (steering magnets, trim quadrupoles, chromaticity and harmonic sextupoles, etc.), ample space for diagnostics, low beta and dispersion functions (to make the beam size small), large dynamic aperture (to accommodate beam halo), and large momentum acceptance (to allow for bunch compression when necessary).

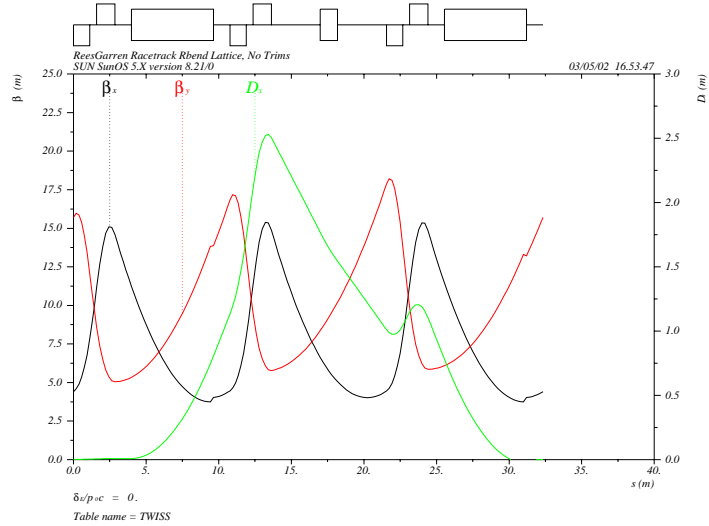

Figure 1. Lattice module of the Fermilab new 8 GeV synchrotron design. Each module has three doublet cells. The dipole in the mid-cell is short. The phase advance per module is 0.8 and 0.6 in the h- and v-plane, respectively. There are five modules in each arc.

### 4.2. *Space Charge*

Amongst various beam physics issues, the space charge is a major concern. It is often the bottleneck limiting the beam intensity in an intense proton source, in particular, in a synchrotron, because the injection energy is low.

A useful scaling factor is the Laslett tune shift

(2)

$$\Delta\nu = -(3r/2) \times (N/\varepsilon_N) \times (1/\beta\gamma^2) \times B_f$$

in which $r$ is the classical proton radius ($1.535 \times 10^{-18}$ m), $N$ the total number of protons, $\varepsilon_N$ the normalized 95% transverse emittance, $\beta$ and $\gamma$ the relativistic factors, and $B_f$ the bunching factor (ratio between peak and average beam current). It shows the space charge effect is most severe at injection because $\beta\gamma^2$ takes the minimum value. The situation becomes worse for high-intensity machines not only because the intensity is high but also because the injection time is long. Numerical simulation is the main tool to study this effect. A number of 1-D, 2-D and 3-D codes have been or are being written at many institutions. An example is shown in Figure 2. These codes are particularly useful to the design of the injection orbit bump current waveform for achieving uniform particle distribution in the beam, reducing emittance dilution and minimizing average number of hits per particle on the stripping foil during the phase space painting process. Several other measures, e.g., tune ramp, inductive inserts, quadrupole mode damper and electron beam compensation are under investigation for possible cures of the space charge effects.

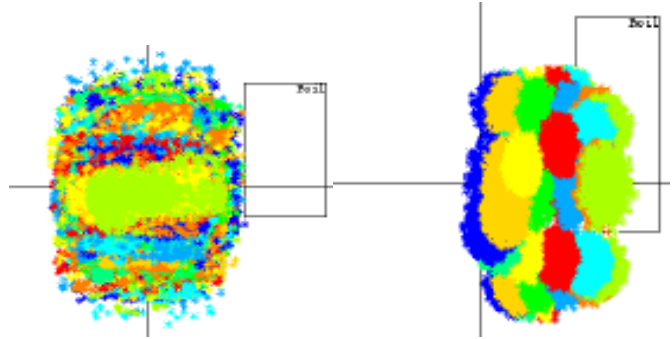

Figure 2. Space charge simulation using Track2D (by C. Prior). It shows the particle distribution after 45 turns injection in the Fermilab new 8 GeV synchrotron with (left) and without (right) the space charge effect.

### 4.3. *Other Beam Dynamics Issues*

In addition to the space charge, there are several other beam dynamics issues that need to be studied concerning an intense proton source.

- Electron cloud effect (ECE). This has been discussed in Section 3.3.5. It is interesting to note that, by far all reported ECE is either in DC machines (accumulators and storage rings) or AC machines in DC operation (i.e., on flat top or flat bottom). No ECE has been seen in AC machines during ramp. Does this imply that AC machines are immune to ECE? If true, this would be an important advantage of the synchrotron approach. However, this is solely an empirical observation. Lack of a reliable theory for understanding and analyzing the ECE is a loophole that urgently needs to be filled.
- Microwave instability of bunched beam below transition. Because the machine will always operate below transition, the negative mass instability due to space charge would not occur. Would then this machine be immune to the microwave instability?
- Tune split. A split between the horizontal and vertical tunes is required in order to avoid the strong resonance $2\nu_x - 2\nu_y = 0$ that could be excited by the space charge. However, it is not clear how big the split needs to be. Does it have to be an integer? Or would a half-integer suffice?

### 4.4. *Beam Loss, Collimation and Remote Handling*

This part is similar to that for accumulators as discussed in Sections 3.3.1 and 3.3.2. There is, however, an important difference. Because the injected linac beam power is low (e.g., 6% of the full beam power if the acceleration range of the synchrotron is 16), higher beam loss at injection (which usually accounts for most of the total loss) can be tolerated. This is an advantage of the synchrotron approach.

### 4.5. *Slow Extraction*

In addition to one-turn extraction, synchrotrons are also used for experiments that require slow extraction (many turns). A critical issue is the efficiency. Although the efficiency of one-turn fast extraction can exceed 99%, it is much lower for multi-turn slow extractions. At high-intensity operation, the beam loss in existing machines during slow spill is usually around 4-5%. This is not acceptable for the next generation of high-intensity machines, in which the beam power will be 1 MW or higher and one percent loss would mean 10 kW or higher. This is a serious problem in the case of KAMI and CKM at the Fermilab Main Injector, and kaon and nuclear physics programs at the JHF. A recent ICFA mini-workshop was devoted to this topic.[10]

**4.6. *Hardware***

4.6.1. *Magnets*

Magnets are one of the most expensive technical systems of a synchrotron. A critical parameter in the magnet design is the vertical aperture of the main bending magnets. The magnet cost is essentially proportional to the aperture. It should be large enough to accommodate a full size beam including its halo. The following criterion can be used in design:

$$A = \{3\, \varepsilon_N \times \beta_{max} / \beta\gamma\}^{1/2} + D_{max} \times \Delta p/p + \text{c.o.d.} \quad (3)$$

in which $A$ is the half aperture, $\varepsilon_N$ the normalized 95% beam emittance, $\beta_{max}$ the maximum beta-function, $D_{max}$ the maximum dispersion, $\Delta p/p$ the relative momentum spread, c.o.d. the closed orbit distortion. The parameter 3 is the estimated size of the beam halo relative to the beam size.

Because this is an AC machine, field tracking between the dipoles and quadrupoles at high field is an important issue. Trim quads or trim coils are needed. The peak dipole field should not exceed 1.5 Tesla. The peak quadrupole gradient is limited by the saturation at the pole root (not pole tip).

The choice of the coil turn number per pole is a tradeoff between the coil AC loss and voltage-to-ground. The former requires the use of many small size coils, whereas the latter requires the opposite, namely, small number of turns. There are two ways to compromise. One is to employ stranded conductor coils, as shown in Figure 3, which was adopted in the JHF 3 GeV ring design. Another is to connect several coils in parallel at the magnet ends, as done in the ISIS. The ratio of the AC *vs*. DC coil loss should be kept around 2-3. The voltage-to-ground should not exceed a few KV.

The aperture and good field region should include a rectangular area (instead of an elliptical area). This is because there will be a significant number of particles residing in the corners of the rectangle.

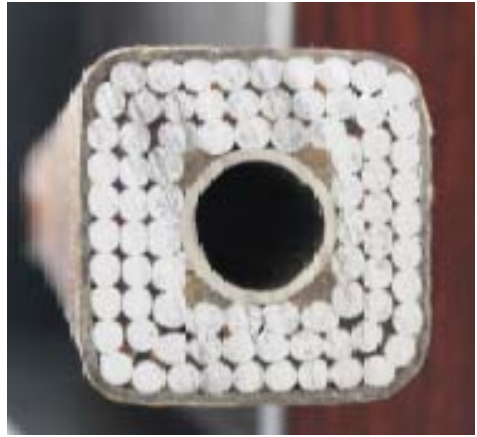

Figure 3. Stranded conductor coil for reducing coil AC loss.

4.6.2. *Power supplies*

This is another expensive technical system. There are several choices for the power supplies in a rapid cycling machine. (1) A single harmonic resonant system, e.g., the Fermilab Booster which resonates at 15 Hz. (2) A dual-harmonic resonant system, e.g., the Fermilab new 8 GeV synchrotron which uses a 15 Hz component plus a 12.5% 30 Hz component as shown below: [11]

$$I(t) = I_0 - I\,cos(2\pi ft) + 0.125\,I\,sin(4\pi ft) \qquad (4)$$

in which f = 15 Hz, $I_0$ and $I$ are two constants determined by the injection and peak current. The advantage of this system is that the peak value of *dB/dt* is decreased by 25%, which leads to a saving of the peak RF power by the same amount. (3) A programmable ramp system, e.g., the AGS Booster and AGS. Although this is a most versatile system (e.g., allowing for a front porch and a flat top), it is also most expensive.

4.6.3. *RF*

The RF system is demanding, because it must deliver a large amount of power to the beam in a short period. In addition, it must be tunable, because the particle revolution frequency increases during acceleration. Cavities with ferrite tuners have been in use for decades. Recently the development of the Finemet cavities at the KEK has aroused strong interest at many laboratories. Thanks to a US-Japan collaboration, Fermilab has built a 7.5 MHz, 15 kV Finemet cavity and installed it in the Main Injector for bunch coalescing. [12] The main advantages of the Finemet cores are high accelerating gradient and wide bandwidth. The former is especially important for high intensity small size rings, in which space is precious. The main concern, however, is its high power consumption. For example, the Fermilab Finemet cavity needs a 200 kW power amplifier to drive it. New types of magnetic alloys are under investigation for performance improvement.

4.6.4. *Vacuum*

Vacuum pipe for a rapid cycling machine is probably one of the most challenging items. Ceramic pipe with a metallic cage inside has been successfully employed at the ISIS. However, this is a costly solution, because it occupies a significant portion of the magnet aperture. Assuming the ceramic wall and the cage need a 1-in vertical space, a 4-in aperture magnet would have to increase its vertical gap by 25% to 5-in in order to accommodate this pipe. This would directly be translated to a 25% increase in the magnet and power supply costs, equivalent to tens of millions dollars.

Thin metallic pipe is an alternative. However, it must be very thin (several mils) in order to minimize the eddy current effects (pipe heating and induced magnetic field). Such a thin pipe is mechanically unstable under vacuum. Several designs have been tried to enhance its stability, including ceramic shields, metallic ribs and spiral lining. Prototyping of the first two designs did not work well. The third one looks promising and is currently under investigation. [13]

Another alternative is that the magnets employ external vacuum skins like those in the Fermilab Booster. Perforated metallic liners are used in the magnet gap to provide a low-impedance environment for the beam.

4.6.5. *Diagnostics*

In addition to the conventional diagnostics for measuring beam position, tune, profile, intensity and loss, intense proton sources have several specific requirements. A system that can diagnose beam parameters during multi-turn injection is highly desirable. The method for fast, accurate non-invasive tune measurement is being developed. A circulating beam profile monitor covering a large dynamic range with turn-by-turn speed will be crucial for studying beam halo. (A similar instrument has been developed for the linac beam halo experiment at Los Alamos. [7]) There was also an ICFA mini-workshop devoted to this topic. [14]

**4.7.** *New Ideas*

In the past several years, there are a number of new or revitalized ideas proposed to the high intensity proton source study. Here are a few examples:

4.7.1. *Inductive inserts*

They are made of ferrite rings and also can have bias current for impedance tuning. Their inductive impedance would fully or partially compensate the space charge impedance, which is capacitive. The first successful experiment was at the PSR. [15] Two ferrite modules made by Fermilab have been installed in the ring. They help increase the e-p instability threshold, which is a major bottleneck of that machine. Another experiment is going on at the Fermilab Booster.

4.7.2. *Induction synchrotron*

This is a longitudinally separated function machine. In other words, the longitudinal focusing and acceleration are carried out by two separate RF systems. The former uses barrier RF buckets, the latter a constant RF voltage curve. One useful feature of this type of machine is tunable bunch lengths. So the so-called superbunch acceleration could be possible. Because a superbunch is similar to a debunched beam, the peak beam current is low. Thus, the space charge effect can be reduced and beam intensity increased.

4.7.3. *Barrier RF stacking*

The application of Finemet and other magnetic alloys makes it possible to build broadband barrier RF cavities with high voltage (~10 kV or higher). They can be used to stack beams in the longitudinal phase space. This is particularly useful when the beam intensity of a synchrotron is limited by its injector (e.g., the intensity of the Fermilab Main Injector is limited by the Booster). Compared to the slip stacking, an advantage of barrier RF stacking is the greatly reduced beam loading effects due to a lower peak beam current. [16,17]

4.7.4. *Fixed field alternating gradient (FFAG) accelerator*

Although MURA first proposed this idea about 40 years ago, it was almost forgotten. Only the recent activities at the KEK brought it back to the world's attention. KEK has successfully built a 1 MeV Proof-of-Principle (PoP) proton FFAG and is building a 150 MeV one. [18] FFAG is an ideal machine for high intensity beams. Its repetition rate can be much higher than a rapid cycling synchrotron (in the range of kHz). One problem of the FFAG, however, is that it is difficult (if not impossible) to fit it into an existing accelerator complex, which usually consists of a linac and a cascade of synchrotrons.

### 4.7.5. *Repetition rate increase in existing synchrotrons*

This is a brute force approach but can be appealing because it is straightforward. For example, the Brookhaven National Laboratory has a proposal for increasing the AGS repetition rate from 0.5 Hz to 2.5 Hz. [19] The Fermilab Main Injector upgrade also includes a rep rate increase (from 0.53 Hz to 0.65 Hz). [20]

## 5. Design Concept of a Proton Driver

### 5.1. *Differences between a Proton Driver and a Spallation Neutron Source*

A proton driver is a high intensity proton source. It can be used as a spallation neutron source. But it can do more. It can generate neutrino superbeams and other high intensity secondary particles (muons, kaons, pions, antiprotons, etc.) for high-energy physics experiments. It can also be used as the first stage of a neutrino factory and a muon collider.

There are two main differences between a proton driver and a spallation neutron source.

- The beam energy of a proton driver is higher. A commonly used production target is carbon. For a carbon target, the $\pi^-$ cross-section is much lower than $\pi^+$ when the proton beam energy is below 4 GeV. Therefore, for polarized muon experiments, a proton driver must be 4 GeV or higher. Furthermore, for neutrino oscillation experiments, a proton source with tunable energy in the range of several GeV up to about 100 GeV is preferred.
- The bunch length of a proton driver is shorter. The pion yield (i.e., number of pions per unit proton beam power) has a strong dependence on the proton bunch length. This is the only parameter that we have control to minimize the 6-D phase space volume of the pions. Moreover, to obtain highly polarized pion beams also requires short proton bunch length. The typical bunch length in a proton driver is a few ns (instead of $\mu$s as in a spallation neutron source).

### 5.2. *How to Achieve Higher Energies*

In a synchrotron-based design, this is not difficult. The energy covers a wide range: from as low as 3 GeV (JHF, 1 MW) to as high as 120 GeV (Fermilab Main Injector upgrade, 2 MW).

In a linac-based design, however, this is severely limited by the cost. The existing highest energy proton linac is the LANSCE (0.8 GeV) at Los Alamos. The SNS linac under construction at Oak Ridge is 1 GeV. There are proposals for 2.2 GeV (CERN), 3 GeV (Los Alamos) and 8 GeV (Fermilab) proton linacs. But none of these has become a construction project.

### 5.3. *How to Obtain Short Bunch Lengths*

In a synchrotron-based design, the bunch length is determined by the RF bucket length, i.e., by the RF frequency. A short bunch length implies the use of a high frequency RF system. However, sometimes there are good reasons to use low frequency RF (e.g., to limit the number of bunches). In this case, a bunch rotation technique can be used for compressing the bunch length.

It should be pointed out that there is a new beam dynamics problem associated with bunch rotation, namely, the path length dependence on momentum spread $\Delta p/p$ and space charge tune shift $\Delta \nu$. This is especially important for proton drivers, in which due to large momentum spread (a few percent) and large tune shift (a few tenth), the dependence of the path length $\Delta L$ on $\Delta p/p$ and $\Delta \nu$ can no longer be ignored. In other words, the momentum compaction factor $\alpha = (\Delta L/L) / (\Delta p/p)$ cannot be treated as a constant during bunch rotation. It is dependent upon the momentum and amplitude of each particle. This will result in a longer bunch after rotation. Simulation study must take this effect into account.

In a linac-based design, a compressor ring (separate from an accumulator ring) will be needed in order to provide the required bunch length and bunch structure.

## 6. Summary

Two recent major spin-offs from high-energy accelerators are synchrotron light sources and high intensity proton sources. Both have found wide-range applications in the field of basic sciences (e.g., material science, molecular biology, chemistry, etc.) as well as in industrial research and development (e.g., chip technology, nano technology, medical and pharmaceutical research, etc.). Spallation neutron source is an important type of the latter.

There are two approaches to a spallation neutron source. One is linac-based, another synchrotron-based. Each approach has its pros and cons. The PSR at the Los Alamos National Laboratory and the SNS project at the Oak Ridge National Laboratory belong to the former, while the ISIS at the Rutherford Appleton Laboratory and the JHF project at the KEK/JAERI represent the latter. (Note that the JHF is a multi-purpose facility unlike the SNS, which serves solely as a neutron source.)

There are close connections between the design of a spallation neutron source and a proton driver. The latter is a strong contender for a near term construction project in the high-energy physics field in the U.S., Europe and Japan. The studies of the two types of machines benefit each other.

The work on high intensity proton sources has been a dynamic field in the accelerator world. There are numerous challenging problems as well as great expectations. Out of the world's three large accelerator projects currently under construction - LHC, SNS and JHF - two are high intensity proton sources. Several more have appeared on the horizon. We'd like to encourage young people to join this field and bring with them their energy, enthusiasm and fresh ideas.

**Acknowledgements**

**Appendix A**

There have been numerous conferences and workshops on high intensity proton sources sponsored by the Beam Dynamics Panel of the International Committee for Future Accelerators (ICFA). For example, there are a series of ICFA mini-workshops on various specific topics, including transition crossing, particle losses, RF, beam loading, transverse and longitudinal emittance measurement and preservation, injection and extraction, beam halo and scraping, two-stream instability, diagnostics and space charge simulations. These mini-workshops can be found on the web http://www-bd.fnal.gov/icfa/workshops/workshops.html. Paper proceedings are also available from the workshop organizers.

There was an ICFA-HB2002 workshop in April 2002 at Fermilab, which covered almost all the aspects concerning high intensity proton sources. The web address is http://www-bd.fnal.gov/HB2002/.

There was an ECLOUD'02 workshop also in April 2002 at CERN for the study of electron cloud effect. The proceedings and presentations are posted at http://wwwslap.cern.ch/collective/ecloud02/.

An international workshop on induction accelerators took place in October 2002 at the KEK. The web address is http://conference.kek.jp/RPIA2002/.

**Appendix B**

In July 2001, about 1,200 physicists over the world gathered at Snowmass, Colorado, USA, for three weeks to discuss the future of high-energy physics. One specific topic was high intensity proton sources. A detailed 26-point R&D program was crafted. This program is directly related to the spallation neutron source work. The Executive Summary is attached, which can be used as guidance for planning future R&D. The full context can be found on the web: http://www-bd.fnal.gov/icfa/snowmass/.

*Executive Summary of Snowmass2001 on High Intensity Proton Sources*

The US high-energy physics program needs an intense proton source (a 1-4 MW Proton Driver) by the end of this decade. This machine will serve multiple purposes: (i) a stand-alone facility that will provide neutrino superbeams and other high intensity secondary beams such as kaons, muons, neutrons, and anti-protons (cf. E1 and E5 group reports); (ii) the first stage of a neutrino factory (cf. M1 group report); (iii) a high brightness source for a VLHC (cf. M4 group report).

Based on present accelerator technology and project construction experience, it is both feasible and cost-effective to construct a 1-4 MW Proton Driver. There are two PD design studies, one at FNAL and the other at the BNL. Both are designed for 1 MW proton beams at a cost of about US$200M (excluding contingency and overhead) and upgradeable to 4 MW. An international collaboration between FNAL, BNL and KEK on high intensity proton facilities addresses a number of key design issues. The sc cavity, cryogenics, and RF controls developed for the SNS can be directly adopted to save R&D efforts, cost, and schedule. PD studies are also actively pursued at Europe and Japan.

There are no showstoppers towards the construction of such a high intensity facility. Key research and development items are listed below ({} indicates present status). Category A indicates items that are not only needed for future machines but also useful for the improvement of existing machine performance;

category B indicates items crucial for future machines and/or currently underway.

1) H⁻ source: Development goals - current 60–70 mA {35 mA}, duty cycle 6–12% {6%}, emittance 0.2 π mm-mrad rms normalized, lifetime > 2 months {20 days}. (A)
2) LEBT chopper: To achieve rise time < 10 ns {50 ns}. (B)
3) Study of 4-rod RFQ at 400 MHz, 100 mA, 99% efficiency, HOM suppressed. (B)
4) MEBT chopper: To achieve rise time < 2 ns {10 ns}. (B)
5) Chopped beam dump: To perform material study & engineering design for dumped beam power > 10 kW. (A)
6) Funneling: To perform (i) one-leg experiment at the RAL by 2006 with goal one-leg current 57 mA; (ii) deflector cavity design for CONCERT. (all B)
7) Linac RF control: To develop (i) high performance HV modulator for long pulsed (>1ms) and CW operation; (ii) high efficiency RF sources (IOT, multi-beam klystron). (all A)
8) Linac sc RF control: Goal - to achieve control of RF phase error < 0.5° and amplitude error <0.5% {presently 1°, 1% for warm linac}. (i) To investigate the choice of RF source (number of cavity per RF source, use of high-power source); (A) (ii) to perform redundancy study for high reliability; (B) (iii) to develop high performance RF control (feedback and feedforward) during normal operation, tuning phases and off-normal operation (missing cavity), including piezo-electric fast feedforward. (A)
9) Space charge: (i) Comparison of simulation code ORBIT with machine data at FNAL Booster and BNL Booster; (ii) to perform 3D ring code bench marking including machine errors, impedance, and space charge (ORNL, BNL, SciDAC, PPPL). (all A)
10) Linac diagnostics: To develop (i) non-invasive (laser wire, ionization, fluorescent-based) beam profile measurement for H⁻;(ii) on-line measurement of beam energy and energy spread using time-of-flight method; (iii) halo monitor especially in sc environment; (iv) longitudinal bunch shape monitor. (all A)
11) SC RF linac: (i) High gradients for intermediate beta (0.5 – 0.8) cavity; (A) (ii) Spoke cavity for low beta (0.17 – 0.34). (B)
12) Transport lines: To develop (i) high efficiency collimation systems; (A) (ii) profile monitor and halo measurement; (A) (iii) energy stabilization by HEBT RF cavity using feedforward to compensate phase-jitter. (B)
13) Halo: (i) To continue LEDA experiment on linac halo and comparison with simulation; (ii) to start halo measurement in rings and comparison with simulation. (all B)

14) Ring lattice: To study higher order dependence of transition energy on momentum spread and tune spread, including space charge effects. (B)
15) Injection and extraction: (i) Development of improved foil (lifetime, efficiency, support); (A) (ii) experiment on the dependence of $H^0$ excited states lifetime on magnetic field and beam energy; (B) (iii) efficiency of slow extraction systems. (A)
16) Electron cloud: (i) Measurements and simulations of the electron cloud generation (comparison of the measurements at CERN and SLAC on the interaction of few eV electrons with accelerator surfaces, investigation of angular dependence of SEY, machine and beam parameter dependence); (A) (ii) determination of electron density in the beam by measuring the tune shift along the bunch train; (A) (iii) theory for bunched beam instability that reliably predicts instability thresholds and growth rates; (A) (iv) investigation of surface treatment and conditioning; (A) (v) study of fast, wide-band, active damping system at the frequency range of 50–800 MHz. (B)
17) Ring beam loss, collimation, protection: (i) Code benchmarking & validation (STRUCT, K2, ORBIT); (A) (ii) engineering design of collimator and beam dump; (A) (iii) experimental study of the efficiency of beam-in-gap cleaning; (A) (iv) bent crystal collimator experiment in the RHIC; (B) (v) collimation with resonance extraction. (B)
18) Ring diagnostics: (i) Whole area of diagnosing beam parameters during multi-turn injection; (ii) circulating beam profile monitor over large dynamic range with turn-by-turn speed; (iii) fast, accurate non-invasive tune measurement. (all A)
19) Ring RF: To develop (i) low frequency (~5 MHz), high gradient (~1 MV/m) burst mode RF systems; (B) (ii) high gradient (50-100 kV/m), low frequency (several MHz) RF system with 50-60% duty cycle; (B) (iii) high-voltage (>100 kV) barrier bucket system; (B) (iv) transient beam loading compensation systems (e.g. for low-Q MA cavity). (A)
20) Ring magnets: (i) To develop stranded conductor coil; (ii) to study voltage-to-ground electrical insulation; (iii) to study dipole/quadrupole tracking error correction. (all B)
21) Ring power supplies: To develop (i) dual-harmonic resonant power supplies; (ii) cost effective programmable power supplies. (all B)
22) Kicker: (i) Development of stacked MOSFET modulator for DARHT and AHF to achieve rise/fall time <10-20 ns; (B) (ii) impedance reduction of lumped ferrite kicker for SNS. (A)
23) Instability & impedance: (i) To establish approaches for improved estimates of thresholds of fast instabilities, both transverse and longitudinal (including space charge and electron cloud effects); (ii) to place currently-used models such as the broadband resonator and distributed impedance on a firmer theoretical basis; (iii) impedance measurement based on coherent tune shifts *vs*. beam intensity, and instability growth rate *vs*. chromaticity, including that for flat vacuum chambers; (iv) to develop new technology in feedback implementation. (all B)
24) FFAG: (i) 3-D modeling of magnetic fields and optimization of magnet profiles; (ii) wide-band RF systems; (iii) transient phase shift in high frequency RF structures; (iv) application of sc magnets. (all B)
25) Inductive inserts: (i) Experiments at the FNAL Booster & JHF3; (A) (ii) programmable inductive inserts; (B) (iii) development of inductive inserts which have large inductive impedance and very small resistive impedance; (B) (iv) theoretical analysis. (B)
26) Induction synchrotron: (i) Study of beam stability; (ii) development of high impedance, low loss magnetic cores. (all B)